\documentclass[12pt, prd, preprint, amsmath, amssymb, aps, superscriptaddress,eqsecnum, floats,nofootinbib,preprintnumbers]{revtex4-1}
\usepackage{color,graphicx,epsfig}
\usepackage{ifpdf}
\usepackage{amsmath}
\usepackage{bm}
\usepackage{color}
\usepackage[english]{babel}
\usepackage{graphicx}%
\usepackage{amsfonts}%
\usepackage{amssymb}
\usepackage{braket}
\usepackage{hyperref}
\usepackage{gensymb}

\definecolor{nicered}{rgb}{0.7,0.1,0.1}
\definecolor{nicegreen}{rgb}{0.1,0.5,0.1}
\hypersetup{colorlinks,citecolor= nicegreen,linkcolor= nicered}

\newenvironment{Eqnarray}{\arraycolsep 0.14em\begin{eqnarray}}{\end{eqnarray}}
\def\beqa{\begin{Eqnarray}}
\def\eeqa{\end{Eqnarray}}
\newcommand{\no}{\nonumber}
\newcommand{\abs}[1]{\lvert#1\rvert}

\newcommand{\beq}{\begin{equation}}
\newcommand{\eeq}{\end{equation}}
\newcommand{\bea}{\begin{eqnarray}}
\newcommand{\eea}{\end{eqnarray}}

\def\lsim{\mathrel{\rlap{\lower4pt\hbox{\hskip1pt$\sim$}}
     \raise1pt\hbox{$<$}}}         
\def\gsim{\mathrel{\rlap{\lower4pt\hbox{\hskip1pt$\sim$}}
     \raise1pt\hbox{$>$}}}         

\newcommand\cmt[1]{}
\definecolor{orchid}{rgb}{0.854902,0.439216,0.839216}

\usepackage[nolist]{acronym}
\begin{acronym}
	\acro{SM}{Standard Model}
	\acro{CPV}{CP violating}
	\acro{CPC}{CP conserving}
	\acro{BSM}{beyond the Standard Model}
	\acro{EFT}{Effective Field Theory}
	\acro{RHS}{right hand side}
	\acro{LHS}{left hand side}
	\acro{CL}{confidence level}
	\acro{GMFV}{General Minimal Flavor Violation}
	\acro{MFV}{Minimal Flavor Violation}
	\acro{FN}{Froggatt-Nielsen}
	\acro{FCNC}{flavor changing neutral current}
\end{acronym}

\begin{document}


\vskip1.5cm
\begin{center}
{\Large \bf Lessons from the LHCb measurement of CP violation in $B_s\to K^+K^-$}
\end{center}
\vskip0.2cm

\begin{center}
Yosef Nir$^1$,  Inbar Savoray$^1$ and Yehonatan Viernik$^1$\\
\end{center}
\vskip 8pt

\begin{center}
{ \it $^1$Department of Particle Physics and Astrophysics,\\
Weizmann Institute of Science, Rehovot 7610001, Israel} \vspace*{0.3cm}

{\tt  yosef.nir,inbar.savoray,yehonatan.viernik@weizmann.ac.il}
\end{center}

\vglue 0.3truecm

  \vskip 3pt \noindent
\centerline{\large Abstract}
The LHCb experiment measured the time-dependent CP asymmetries $C_{KK}$ and $S_{KK}$ in $B_s\to K^+K^-$ decay. Combining with the corresponding CP asymmetries $C_{\pi\pi}$ and $S_{\pi\pi}$ in $B\to \pi^+\pi^-$ decay, we find that the size of $U$-spin breaking in this system is of order $20\%$. Moreover, the data suggest that these effects are dominated by factorizable contributions. We further study the constraints on new physics contributions to $b\to u\bar uq$ ($q=s,d$). New physics that is minimally flavor violating (MFV) cannot be distinguished from the \ac{SM} in these decays. However, new physics that is not MFV can mimic large $U$-spin breaking. Requiring that the $U$-spin breaking parameters remain below the size implied by the data leads to a lower bound of $5-10$ TeV on the scale of generic new physics. If the new physics is subject to the selection rules that follow from the \ac{FN} mechanism or from \ac{GMFV}, the bound is relaxed to 2 TeV. 

\newpage
\acresetall
\section{Introduction}
CP violation in neutral meson decays has provided stringent tests of the \ac{SM} and has been a very effective probe of new physics, with examples such as the measurement of CP violation in $K_L\to\pi\pi$ decays \cite{Christenson:1964fg} which led to the prediction that there is a third generation of fermions \cite{Kobayashi:1973fv}, and the measurement of time-dependent CP violation in $B\to J/\psi K_S$ decays \cite{Aubert:2001nu,Abe:2001xe} which proved that the Kobayashi-Maskawa mechanism is the dominant source of the observed CP violation, and excluded alternatives such as the superweak CP violation model \cite{Wolfenstein:1974fd} and approximate CP \cite{Eyal:1998bk}. For a recent review, see \cite{Nir:2020mgy}.

Recently, the LHCb collaboration provided the first observation of time-dependent CP violation in $B_s$ decays \cite{LHCb:2020byh}.
The time dependent CP asymmetry in $B_s\to K^+K^-$ decay is given by
\beqa
{\cal A}_{KK}^s(t)&\equiv&\frac{\Gamma_{\overline{B_s}\to K^+K^-}(t)- \Gamma_{{B_s}\to K^+K^-}(t)}
{\Gamma_{\overline{B_s}\to K^+K^-}(t)+\Gamma_{{B_s}\to K^+K^-}(t)}\\
&=&\frac{-C_{KK}\cos(\Delta m_st)+S_{KK}\sin(\Delta m_s t)}
{\cosh(\Delta\Gamma_st/2)+A_{KK}^{\Delta\Gamma}\sinh(\Delta\Gamma_st/2)}.\no
\eeqa
CP symmetry would imply $C_{KK}=S_{KK}=0$ and $A^{\Delta\Gamma}_{KK}=1$. Combining the measurement reported in Ref. \cite{LHCb:2020byh} with the one reported in Ref. \cite{Aaij:2018tfw}, the ranges of the relevant parameters are
\beqa\label{eq:csaexp}
C_{KK}&=&+0.172\pm0.031,\no\\
S_{KK}&=&+0.139\pm0.032,\no\\
A_{KK}^{\Delta\Gamma}&=&-0.897\pm0.087.
\eeqa
In this work we present the theoretical interpretation of this measurement, and explore what can (or cannot) be learned from it.

Given that there are hadronic parameters playing a role, one can go in two directions:
\begin{itemize}
\item Assume the \ac{SM}, and extract the hadronic parameters. Then we can learn about $U$-spin breaking by comparing to the $U$-spin related parameters extracted from $B_d\to\pi^+\pi^-$.
\item Use the consistency of the various measurements with approximate $U$-spin symmetry to obtain constraints on new physics contributions to the $b\to u\bar us$  and $b\to u\bar ud$ decays.
\end{itemize}

For the latter study, we assume that, for processes that get contributions from \ac{SM} tree-level diagrams that are not CKM suppressed, the contributions from new physics can be neglected. We allow, however, contributions of a-priori arbitrary size and phase to \ac{FCNC} processes and to CKM suppressed tree level processes. Concretely for $B_s$ physics, we assume that new physics contributions to $B_s\to J/\psi\phi$ are negligible, while the size and the phase of new physics contributions to $B_s-\overline{B_s}$ mixing and to $B_s\to K^+K^-$ are only constrained by experimental data.

As concerns the assumption that $b\to c\bar cs$ decays are dominated by the \ac{SM}, note that $A(b\to c\bar cs)\propto G_F V_{cb}$. All $B$ and $B_s$ decay amplitudes (and, in particular, the leading semileptonic decays) are suppressed by at least such a factor, or even stronger. If new physics contributes significantly to this decay, then the whole consistency of the data with the CKM picture seems accidental, which is very unlikely. Moreover, for new physics to be comparable to the \ac{SM} contribution, it needs to be lighter than ${\cal O}({\rm TeV})$ and have tree level flavor changing couplings to quarks. Thus, very likely it should have been directly observed at the LHC, and affect other flavor observables significantly. In contrast, the $b\to u\bar us$ decay can get significant contributions from new physics at the 10 TeV scale, and it is Cabibbo-suppressed compared to the charmless semileptonic $b$ decays. The $b\to u\bar ud$ decay is an intermediate case, as it has the same suppression as the charmless semileptonic $b$ decays, and can get significant contributions from new physics lighter than ${\cal O}(5\ {\rm TeV})$.

Much -- if not all -- of our ability to extract lessons on the \ac{SM} and on new physics from $B_s\to K^+K^-$ relies  on the $U$-spin relation with $B_d\to\pi^+\pi^-$. Thus, in the following sections, we analyze side-by-side the two processes. In Section \ref{sec:formalism} we introduce the necessary formalism and notations. In Section \ref{sec:modind} we present model independent considerations that make our analysis as generic and as data-driven as possible. In Section \ref{sec:sm} we use the experimental data to extract the relevant hadronic parameters, and in Section \ref{sec:uspin} we find the size of $U$-spin breaking effects. In Section \ref{sec:np} we put bounds on the size of new physics contributions to the $B_s\to K^+K^-$ and to $B_d\to\pi^+\pi^-$ decays.

\section{Formalism and notations}
\label{sec:formalism}
We follow the formalism and notations presented in Ref. \cite{Zyla:2020zbs}.

\subsection{$B_s\to K^+K^-$}
The neutral mass eigenstates of the $B_s-\overline{B_s}$ system are given by
\beq\label{eq:psqs}
|B_{sL,sH}\rangle=p_s|B_s\rangle\pm q_s|\overline{B_s}\rangle,
\eeq
with the normalization $|p_s|^2+|q_s|^2=1$. In terms of the dispersive and absorptive parts of the $\Delta B=2$ transition amplitudes, we have
\beq
\left(\frac{q_s}{p_s}\right)^2=\frac{M^*_{B_s\overline{B_s}}-(i/2)\Gamma^*_{B_s\overline{B_s}}}
{M_{B_s\overline{B_s}}-(i/2)\Gamma_{B_s\overline{B_s}}}.
\eeq

We define the decay amplitudes,
\beq\label{eq:afbaraf}
A^s_{f}=\langle f|{\cal H}|B_s\rangle,\ \ \ \overline{A}^s_{\bar f}=\langle \overline{f}|{\cal H}|\overline{B_s}\rangle,
\eeq
and the parameter $\lambda_{f}^s$:
\beq\label{eq:lambdaf}
\lambda_{f}^s\equiv\frac{q_s}{p_s}\frac{\overline{A}^s_{f}}{{A}^s_{f}}.
\eeq

We denote $A_{KK}\equiv A^s_{K^+K^-}$ and $\lambda_{KK}\equiv\lambda^s_{KK}$. The parameters of Eq. (\ref{eq:csaexp}) are given by
\beq\label{eq:csalambda}
C_{KK}=\frac{1-|\lambda_{KK}|^2}{1+|\lambda_{KK}|^2},\ \ \
S_{KK}=\frac{2{\cal I}m\lambda_{KK}}{1+|\lambda_{KK}|^2},\ \ \
A_{KK}^{\Delta\Gamma}=\frac{-2{\cal R}e\lambda_{KK}}{1+|\lambda_{KK}|^2}.
\eeq
Note that a consistency check is provided by $(C_{KK})^2+(S_{KK})^2+(A_{KK}^{\Delta\Gamma})^2=1$.

The $B_s\to K^+K^-$ decay goes via the $b\to u\bar us$ quark transition. It depends on the following CKM combinations:
\beq
\lambda^j_{bs}\equiv V_{jb}^*V_{js}\ \ \ (j=u,c,t).
\eeq
Within the \ac{SM}, one can write the decay amplitudes as follows:
\beqa\label{eq:akk}
A_{KK}^{\rm SM}&=&\hat P^s\lambda^c_{bs}+\hat T^s\lambda^u_{bs}
=\hat P^s \lambda^c_{bs}(1+r_s\lambda^u_{bs}/\lambda^c_{bs}),\no\\
\overline{A}_{KK}^{\rm SM}&=&\hat P^s\lambda^{c*}_{bs}+\hat T_s\lambda^{u*}_{bs}
=\hat P^s \lambda^{c*}_{bs}(1+r_s\lambda^{u*}_{bs}/\lambda^{c*}_{bs}).
\eeqa
A few comments are in place regarding Eqs. (\ref{eq:akk}):
\begin{itemize}
\item Within the \ac{SM}, the only source of CP violation is the CKM matrix. Hence, only the $\lambda^j_{bs}$ factors are complex conjugated between $A_{KK}^{\rm SM}$ and $\overline{A}_{KK}^{\rm SM}$.
\item The $r_s$ factor is, in general, complex. Its phase is a so-called strong phase, which is the same in $A_{KK}$ and $\overline{A}_{KK}$.
\item   $\hat P^s$, $\hat T^s$ (and $r_s$) include in them not only QCD matrix elements, but also electroweak parameters which are neither flavor dependent nor CP violating, such as $G_F$.
\item The $b\to u\bar us$ transition has \ac{SM} tree and penguin contributions. We define $T^s$ as the tree contribution, and $P^s_j$ as the penguin contribution with an intermediate $j$-quark. Then, using CKM unitarity, we have
\beq\label{eq:tsps}
\hat T^s=T^s+P^s_u-P^s_t,\ \ \ \hat P^s=P^s_c-P^s_t,\ \ \  r_s\equiv \hat T^s/\hat P^s.
\eeq
\end{itemize}

\subsection{$B_d\to \pi^+\pi^-$}
We define $q_d$ and $p_d$ for the $B_d-\overline{B_d}$ system in a similar way to Eq. (\ref{eq:psqs}), and $A^d_f$, $\overline{A}^d_f$ and $\lambda^d_f$ in a similar way to Eqs. (\ref{eq:afbaraf}) and (\ref{eq:lambdaf}). We denote $A_{\pi\pi}\equiv A^d_{\pi^+\pi^-}$ and $\lambda_{\pi\pi}\equiv\lambda^d_{\pi\pi}$. The parameters of the time-dependent CP asymmetry in $B\to\pi^+\pi^-$ are given by
\beq\label{eq:csalambdad}
C_{\pi\pi}=\frac{1-|\lambda_{\pi\pi}|^2}{1+|\lambda_{\pi\pi}|^2},\ \ \
S_{\pi\pi}=\frac{2{\cal I}m\lambda_{\pi\pi}}{1+|\lambda_{\pi\pi}|^2},\ \ \
A_{\pi\pi}^{\Delta\Gamma}=\frac{-2{\cal R}e\lambda_{\pi\pi}}{1+|\lambda_{\pi\pi}|^2}.
\eeq

The $B_d\to \pi^+\pi^-$ decay goes via the $b\to u\bar ud$ quark transition. It depends on the following CKM combinations:
\beq
\lambda^j_{bd}\equiv V_{jb}^*V_{jd}\ \ \ (j=u,c,t).
\eeq
Within the \ac{SM}, one can write the decay amplitudes as follows:
\beqa\label{eq:app}
A_{\pi\pi}^{\rm SM}&=&\hat P^d\lambda^c_{bd}+\hat T^d\lambda^u_{bd}=\hat P^d \lambda^c_{bd}(1+r_d\lambda^u_{bd}/\lambda^c_{bd}),\no\\
\overline{A}_{\pi\pi}^{\rm SM}&=&\hat P^d\lambda^{c*}_{bd}+\hat T^d\lambda^{u*}_{bd}=\hat P^d \lambda^{c*}_{bd}(1+r_d\lambda^{u*}_{bd}/\lambda^{c*}_{bd}),
\eeqa
where $r_d$ is, in general, complex. The $b\to u\bar ud$ transition has \ac{SM} tree and penguin contributions.
We define $T^d$ as the tree contribution, and $P^d_j$ as the penguin contribution with intermediate $j$-quark. Then, using CKM unitarity, we have
\beq\label{eq:tdpd}
\hat T^d=T^d+P^d_u-P^d_t,\ \ \ \hat P^d=P^d_c-P^d_t,\ \ \  r_d\equiv \hat T^d/\hat P^d.
\eeq
Note again that $\hat P^d$, $\hat T^d$ (and $r_d$) include in them not only QCD matrix elements, but also flavor-universal CP conserving electroweak parameters.

\section{Model-independent considerations}
\label{sec:modind}
\subsection{$B_s\to K^+K^-$}
The CP asymmetry in wrong-sign semileptonic $B_s$ decays is given by
\beqa
{\cal A}_{\rm SL}^s&\equiv&\frac{d\Gamma/dt[\overline{B_s}(t)\to\ell^+X]-d\Gamma/dt[B_s(t)\to\ell^-X]}
{d\Gamma/dt[\overline{B_s}(t)\to\ell^+X]+d\Gamma/dt[B_s(t)\to\ell^-X]}\\
&=&\frac{1-|q_s/p_s|^4}{1+|q_s/p_s|^4}.
\eeqa
%
%
%
The experimental world average of ${\cal A}_{\rm SL}^s$ is given by \cite{Amhis:2019ckw}
\beq
{\cal A}_{\rm SL}^s=(-0.6\pm2.8)\times10^{-3},
\eeq
implying
\beq
|q_s/p_s|=1.0003\pm0.0014.
\eeq
For our purposes we can then approximate $|q_s/p_s|=1$ and use
\beq\label{eq:qpbs}
\frac{q_s}{p_s}=\frac{M^*_{B_s\overline{B_s}}}{|M_{B_s\overline{B_s}}|}.
\eeq

The time dependent CP asymmetry in $B_s\to J/\psi\phi$ decay is given by
\beqa
{\cal A}_{\psi\phi}^s(t)&\equiv&\frac{\Gamma_{\overline{B_s}\to J/\psi\phi}(t)- \Gamma_{{B_s}\to J/\psi\phi}(t)}
{\Gamma_{\overline{B_s}\to J/\psi\phi}(t)+\Gamma_{{B_s}\to J/\psi\phi}(t)}\\
&=&\frac{-C_{\psi\phi}\cos(\Delta m_st)+S_{\psi\phi}\sin(\Delta m_s t)}
{\cosh(\Delta\Gamma_st/2)+A_{\psi\phi}^{\Delta\Gamma}\sinh(\Delta\Gamma_st/2)}.\no
\eeqa
The $C_{\psi\phi}$, $S_{\psi\phi}$ and $A^{\Delta\Gamma}_{\psi\phi}$ parameters depend on $\lambda_{\psi\phi}$ in a way similar to Eq. (\ref{eq:csalambda}).

Within the \ac{SM}, the $b\to c\bar cs$ transition has tree and penguin contributions. In a way similar to our analysis of $B_s\to K^+K^-$, we can write
\beq\label{eq:apsiphi}
A_{\psi\phi}=\hat T^s_{\psi\phi} \lambda^c_{bs}(1+r_{\psi\phi}\lambda^u_{bs}/\lambda^c_{bs}).
\eeq
Here, however, the second term in parenthesis is both CKM and loop suppressed, and can thus be neglected. We further assume that new physics effects on decay processes with \ac{SM} tree level contributions that are not CKM suppressed are negligible. We thus obtain:
\beq\label{eq:aapsiphi}
\frac{\overline{A}_{\psi\phi}}{A_{\psi\phi}}\simeq\frac{\lambda^{c*}_{bs}}{\lambda^{c}_{bs}}.
\eeq
Eqs. (\ref{eq:aapsiphi}) and  (\ref{eq:qpbs}) imply that $\lambda_{\psi\phi}$ is a pure phase:
\beq\label{eq:lampsiphi}
\lambda_{\psi\phi}=\frac{M^*_{B_s\overline{B_s}}}{|M_{B_s\overline{B_s}}|}
\frac{\lambda^{c*}_{bs}}{\lambda^{c}_{bs}}
\equiv e^{i\phi_s}.
\eeq
Measurements of various CP asymmetries in $B_s$ decays via $b\to c\bar cs$ lead to the following world average for $\phi_s$ \cite{Amhis:2019ckw}:
\beq
\phi_s=0.051\pm0.023.
\eeq

Eqs. (\ref{eq:lampsiphi}) and (\ref{eq:qpbs}) allow us to express $\lambda_{KK}$ as follows:
\beq
\lambda_{KK}=\frac{M^*_{B_s\overline{B_s}}}{|M_{B_s\overline{B_s}}|}\frac{\overline{A}_{KK}}{A_{KK}}
=e^{i\phi_s}\frac{\lambda^{c}_{bs}}{\lambda^{c*}_{bs}}\frac{\overline{A}_{KK}}{A_{KK}}.
\eeq
Using Eq. (\ref{eq:csaexp}), we obtain
\beqa\label{eq:abslamimlam}
|\lambda_{KK}|^2&=&|\overline{A}_{KK}/{A}_{KK}|^2=0.71\pm0.05,\no\\
{\cal I}m\lambda_{KK}&=&
\cos\phi_s{\cal I}m\left(\frac{\lambda^{c}_{bs}}{\lambda^{c*}_{bs}}\frac{\overline{A}_{KK}}{A_{KK}}\right)
+\sin\phi_s{\cal R}e\left(\frac{\lambda^{c}_{bs}}{\lambda^{c*}_{bs}}\frac{\overline{A}_{KK}}{A_{KK}}\right)=0.119\pm0.028.
\eeqa
%

\subsection{$B_d\to\pi^+\pi^-$}
The experimental world average of ${\cal A}_{\rm SL}^d$ is given by \cite{Zyla:2020zbs}
\beq
{\cal A}_{\rm SL}^d=(-2.0\pm1.6)\times10^{-3},
\eeq
implying
\beq
|q_d/p_d|=1.0010\pm0.0008.
\eeq
For our purposes we can then approximate $|q_d/p_d|=1$ and use
\beq\label{eq:qpbd}
\frac{q_d}{p_d}=\frac{M^*_{B_d\overline{B_d}}}{|M_{B_d\overline{B_d}}|}.
\eeq

Similarly to $B_s\to J/\psi \phi$, the $B\to J/\psi K_S$ decay is dominated by a single CKM combination and, furthermore, new physics contributions can be safely assumed to be negligible. Consequently,
\beq\label{eq:aapsiks}
\frac{\overline{A}_{\psi K_S}}{A_{\psi K_S}}\simeq\frac{\lambda^{c*}_{bd}}{\lambda^{c}_{bd}}.
\eeq
Eqs. (\ref{eq:aapsiks}) and  (\ref{eq:qpbd}) imply that $\lambda_{\psi K_S}$ is a pure phase:
\beq\label{eq:lampsiks}
\lambda_{\psi K_S}=\frac{M^*_{B_d\overline{B_d}}}{|M_{B_d\overline{B_d}}|}
\frac{\lambda^{c*}_{bd}}{\lambda^{c}_{bd}}
\equiv e^{i\phi_d}.
\eeq
Measurements of the CP asymmetries in $B_d\to J/\psi K_S$ lead to the following world average for $\phi_d$ \cite{Zyla:2020zbs,Barel:2020jvf}:
\beq
\phi_d=-0.768\pm0.026.
\eeq

Eqs. (\ref{eq:lampsiks}) and (\ref{eq:qpbd}) allow us to express $\lambda_{\pi\pi}$ as follows:
\beq
\lambda_{\pi\pi}=\frac{M^*_{B_d\overline{B_d}}}{|M_{B_d\overline{B_d}}|}\frac{\overline{A}_{\pi\pi}}{A_{\pi\pi}}
=e^{i\phi_d}\frac{\lambda^{c}_{bd}}{\lambda^{c*}_{bd}}\frac{\overline{A}_{\pi\pi}}{A_{\pi}}.
\eeq
Using the experimental values \cite{Zyla:2020zbs}
\beqa\label{eq:csaexpd}
C_{\pi\pi}&=&-0.32\pm0.04,\no\\
S_{\pi\pi}&=&-0.65\pm0.04,
\eeqa
we obtain
\beqa\label{eq:abslamimlamd}
|\lambda_{\pi\pi}|^2&=&|\overline{A}_{\pi\pi}/{A}_{\pi\pi}|^2=1.94\pm0.17,\no\\
{\cal I}m\lambda_{\pi\pi}&=&
\cos\phi_d{\cal I}m\left(\frac{\lambda^{c}_{bd}}{\lambda^{c*}_{bd}}\frac{\overline{A}_{\pi\pi}}{A_{\pi\pi}}\right)
+\sin\phi_d{\cal R}e\left(\frac{\lambda^{c}_{bd}}{\lambda^{c*}_{bd}}\frac{\overline{A}_{\pi\pi}}{A_{\pi\pi}}\right)=-0.96\pm0.06.
\eeqa
%

\section{The Standard Model}
\label{sec:sm}
The values of the CKM parameters that play a role in the $B_s\to K^+K^-$ and $B_d\to\pi^+\pi^-$ decays are known from tree level decays \cite{Zyla:2020zbs,LHCb:2020kho}:
\beqa
R_{uc}^{bs}&\equiv&\left|\lambda^u_{bs}/\lambda^c_{bs}\right|=0.021\pm0.001,\nonumber\\
R_{uc}^{bd}&\equiv&\left|\lambda^u_{bd}/\lambda^c_{bd}\right|=0.410\pm0.025,\nonumber\\
\gamma&\equiv&\arg\left(-\lambda^u_{bd}/\lambda^c_{bd}\right)=(67\pm4)\degree.
\eeqa
Note that, using CKM unitarity relations, we obtain
\beq
\gamma=\arg\left(\lambda^u_{bs}/\lambda^c_{bs}\right)+{\cal O}(\lambda^4),
\eeq
Neglecting the ${\cal O}(\lambda^4)$ correction,we can rewrite the decay amplitudes,
\beqa\label{eq:AkkpipiU}
A_{KK}&=&\hat P^s \lambda^c_{bs}\left[1+r_{s}R^{bs}_{uc}e^{+i\gamma}\right],\no\\
A_{\pi\pi}&=&\hat P^d\lambda^c_{bd}\left[1-r_{d}R^{bd}_{uc}e^{+i\gamma}\right],
\eeqa
and the $\lambda_f$ parameters,
\beqa\label{eq:lamkkpipiU}
\lambda_{KK}&=&e^{i\phi_s}\left[\frac{1+r_{s}R^{bs}_{uc}e^{-i\gamma}}
{1+r_{s}R^{bs}_{uc}e^{+i\gamma}}\right],\no\\
\lambda_{\pi\pi}&=&e^{i\phi_d}\left[\frac{1-r_{d}R^{bd}_{uc}e^{-i\gamma}}
{1-r_{d}R^{bd}_{uc}e^{+i\gamma}}\right].
\eeqa
Defining
\beq
\delta_q\equiv{\rm arg}(r_q),
\eeq
we obtain, for the CP asymmetries,
\beqa
C_{KK}&=&\frac{-2|r_s| R^{bs}_{uc}\sin\delta_s\sin\gamma}{1+2|r_s|R^{bs}_{uc}\cos\delta_s\cos\gamma+(|r_s|R^{bs}_{uc})^2},\no\\
C_{\pi\pi}&=&\frac{2|r_d| R^{bd}_{uc}\sin\delta_d\sin\gamma}{1-2|r_d|R^{bd}_{uc}\cos\delta_d\cos\gamma+(|r_d|R^{bd}_{uc})^2},
\eeqa
\beqa
S_{KK}&=&\frac{\sin\phi_s+2|r_s| R^{bs}_{uc}\cos\delta_s\sin(\phi_s-\gamma)+(|r_s|R^{bs}_{uc})^2\sin(\phi_s-2\gamma)}{1+2|r_s|R^{bs}_{uc}\cos\delta_s\cos\gamma+(|r_s|R^{bs}_{uc})^2},\no\\
S_{\pi\pi}&=&\frac{\sin\phi_d-2|r_d| R^{bd}_{uc}\cos\delta_d\sin(\phi_d-\gamma)+(|r_d|R^{bd}_{uc})^2\sin(\phi_d-2\gamma)}{1-2|r_d|R^{bd}_{uc}\cos\delta_d\cos\gamma+(|r_d|R^{bd}_{uc})^2},
\eeqa
and for the decay rates (averaged over $B_{q}$ and $\overline{B_{q}}$),
\beqa\label{eq:ratesratio}
\Gamma(B_s\to K^+K^-)&=&
|\hat P^s|^2|\lambda^c_{bs}|^2\left[1+(|r_{s}|R_{uc}^{bs})^2+2|r_{s}|R_{uc}^{bs}\cos\delta_{s}\cos\gamma\right],\no\\
\Gamma(B_d\to\pi^+\pi^-)&=&
|\hat P^d|^2|\lambda^c_{bd}|^2\left[1+(|r_{d}|R_{uc}^{bd})^2-2|r_{d}|R_{uc}^{bd}\cos\delta_{d}\cos\gamma\right],
\eeqa

\begin{figure}[t]
 \begin{center}
 \includegraphics[scale=1]{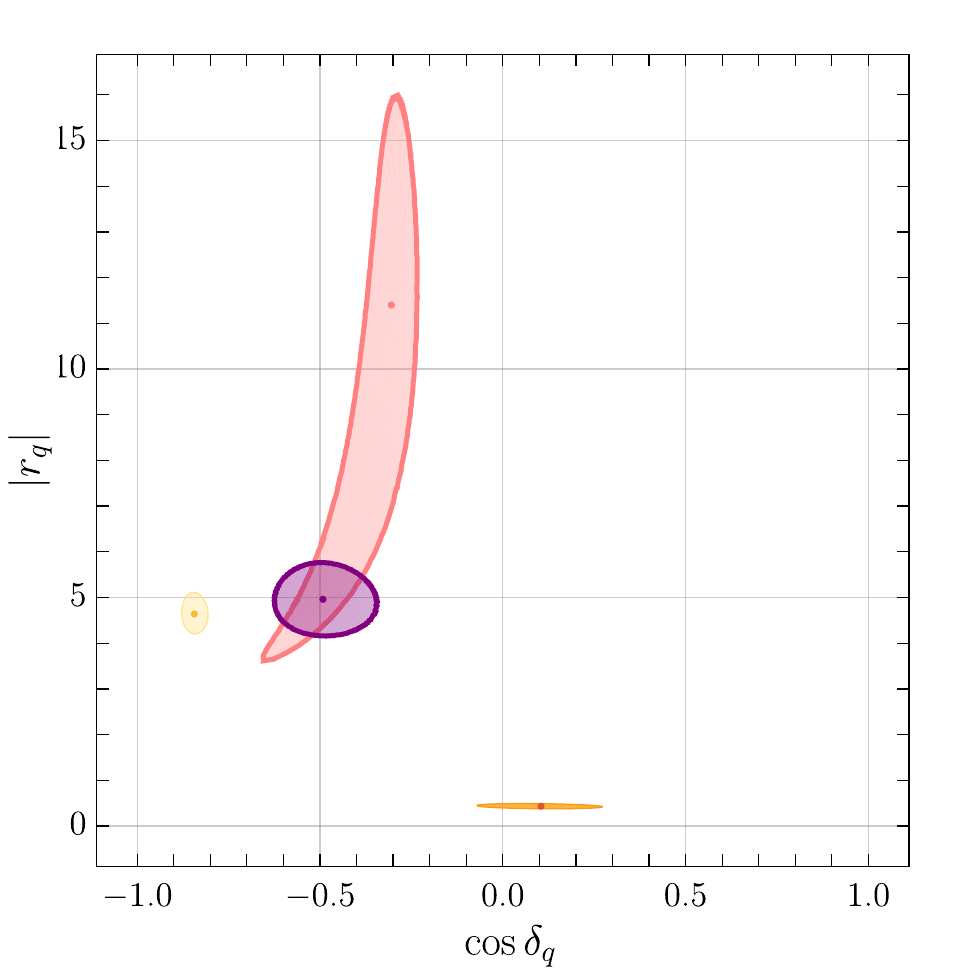} 
 \end{center}
 \caption{The $1\sigma$ allowed ranges for $\abs{r_q}$ vs. $\cos\delta_q$: (purple) $q=s$, from $S_{KK}$ and $C_{KK}$; (pink) $q=s$, from $A^{\Delta\Gamma}_{KK}$ and $C_{KK}$; (yellow) $q=d$, from $S_{\pi\pi}$ and $C_{\pi\pi}$, for negative $A_{\pi\pi}^{\Delta\Gamma}$; (orange)  $q=d$, from $S_{\pi\pi}$ and $C_{\pi\pi}$, for positive $A_{\pi\pi}^{\Delta\Gamma}$.}
\label{fig:rcosdelta}
\end{figure}

Using the experimental values of the five observables $C_{KK}$, $S_{KK}$, $C_{\pi\pi}$, $S_{\pi\pi}$ and
\beq
R_{\Gamma}\equiv\frac{\Gamma(B_s\to K^+K^-)}{\Gamma(B_d\to\pi^+\pi^-)}
=\frac{{\rm BR}(B_s\to K^+K^-)}{{\rm BR}(B_d\to\pi^+\pi^-)}\frac{\tau_d}{\tau_s}=5.2\pm0.5,
\eeq
we can obtain the values of the five hadronic parameters. Solving for the central values of the experimental observables, we obtain:
\beqa\label{eq:exprsrdpspd}
&&|r_s|=4.96 \pm 0.80\,,\ \ \ \cos\delta_s=-0.49 \pm 0.14\ (\sin\delta_s<0)\,,\nonumber\\
&&|r_d|=4.64 \pm 0.45\,,\ \ \ \cos\delta_d=-0.84 \pm 0.04\ (\sin\delta_d<0)\,,\nonumber\\
&&|\hat P_s/\hat P_d|=1.26 \pm 0.11.
\eeqa
Note that, since $A^{\Delta\Gamma}_{\pi\pi}$ is not measured, there is a discrete ambiguity in $|r_d|$ and $\cos\delta_d$. We present the solution that corresponds to a negative  $A^{\Delta\Gamma}_{\pi\pi}$, which gives $\abs{r_d}$ close to $\abs{r_s}$. This is not the case if  $A^{\Delta\Gamma}_{\pi\pi}$ is positive ($|r_d|=0.43$, $\cos\delta_d=+0.10$). Once $|r_q|$ and $\cos\delta_q$ are fixed, the sign of $\delta_{s}$ and $\delta_{d}$ are determined from the sign of $C_{KK}$ and $C_{\pi\pi}$, respectively, as they are the only observables explicitly sensitive to $\sin\delta_q$\,. 
The $1\sigma$ allowed ranges in the $\abs{r_q}-\cos\delta_q$ plane are shown in Fig. \ref{fig:rcosdelta}.

A lesson to be drawn from these results is that, while in the $B_d\to\pi^+\pi^-$ decay the $\hat T^d$ and $\hat P^d$ contributions are comparable, $|r_d|R^{bd}_{uc}\approx1.9$, the $B_s\to K^+K^-$ decay is dominated by the $\hat P^s$ contribution, $|r_s|R^{bs}_{uc}\approx 0.10$.

The fact that $|r_s/r_d|$ and $|\hat P^s/\hat P^d|$ are close to one, suggests that $U$-spin breaking effects are small. This result leads us to investigate in detail the $U$-spin breaking in this system.

\section{$U$-spin breaking}
\label{sec:uspin}
The $B_s\to K^+K^-$ and $B_d\to\pi^+\pi^-$ are related by $U$-spin, the $SU(2)$ symmetry under which $s$ and $d$ form a doublet. Concretely, $U$-spin requires
\beq\label{eq:uspinc}
P\equiv\hat P^s=\hat P^d,\ \ \ T\equiv\hat T^s=\hat T^d.
\eeq
$U$-spin breaking effects are expected to be of ${\cal O}(m_s/\Lambda_{\rm QCD})\sim0.3$. Indeed, the best fit values that we obtained, $|\hat P^s/\hat P^d|=1.26$ and $|r_s/r_d|=1.07$ are consistent with this expectation.

It is interesting to note that the breaking effect in $|\hat P^s/\hat P^d|$ is much larger than the one in $|r_s/r_d|$. The latter is, in fact, a double ratio:
\beq
\frac{r_s}{r_d}=\frac{\hat T_s/\hat T_d}{\hat P_s/\hat P_d}.
\eeq
Factorizable contributions cancel to a good approximation (roughly, $m_s/m_b$) in the double ratio, so the deviation from unity is affected mainly by non-factorizable contributions \cite{Fleischer:1999pa,Ali:2007ff,Beneke:2003zw,He:2017fln}. The data support the assumption that the non-factorizable contributions are small.

To incorporate first-order $U$-spin breaking, we write
\beqa
A_{KK}&=&(P+p)\lambda^c_{bs}+(T+t)\lambda^u_{bs},\no\\
A_{\pi\pi}&=&(P-p)\lambda^c_{bd}+(T-t)\lambda^u_{bd},
\eeqa
where we assume small breaking, {\it i.e.} $|p/P|\ll1$ and $|t/T|\ll1$. Without loss of generality, we can choose $P$ to be real. Then there are seven hadronic parameters.
Given the five observables that we use,  $C_{KK}$, $S_{KK}$, $C_{\pi\pi}$, $S_{\pi\pi}$ and $R_{\Gamma}$, we can extract five of these seven parameters. In principle, we can extract $P$ by considering the individual decay rates, rather than their ratio, but this has no significance for our analysis.

When we consider the $U$-spin breaking effects to first order only, the list of five parameters consists of two $U$-spin conserving ones,
\beqa
\hat T_r&\equiv&{\cal R}e(T/P)=-3.0 \pm 0.5,\no\\
\hat T_i&\equiv&{\cal I}m(T/P)=\pm3.8 \pm 0.6,
\eeqa
and three $U$-spin breaking ones,
\beqa
\hat t_r&\equiv&{\cal R}e(t/T)=+0.14 \pm 0.07,\no\\
\hat p_r&\equiv&{\cal R}e(p/P)=+0.11 \pm 0.04,\no\\
\hat\Delta_i&\equiv&{\cal I}m(t/T-p/P)=\mp0.21 \pm 0.08.
\eeqa
The fourth  $U$-spin breaking parameter, $\hat\Sigma_i\equiv{\cal I}m(t/T+p/P)$, does not play a role at first order in these observables.
We learn that the $U$-spin breaking effects are of order $0.1-0.2$, so that $U$-spin is a good approximate symmetry of this system.

We note that
\beq
\frac{\hat t_r-\hat p_r}{\hat t_r+\hat p_r}\approx0.12\ll1.
\eeq
This is another way of observing that the $U$-spin breaking effects in $\hat P^s/\hat P^d$ and much larger than those in $r_s/r_d$, consistent with the assumption that the leading correction comes from the $f_K/f_\pi$ factor in the factorizable contributions, and that the non-factorizable contributions are small.

A particularly interesting combination of parameters is the following:
\beq\label{eq:defrcg}
R_{C\Gamma}\equiv\frac{C_{KK}\Gamma(B_s\to K^+K^-)}{C_{\pi\pi}\Gamma(B_d\to\pi^+\pi^-)}.
\eeq
It has been noted by Gronau, that in the $U$-spin limit, the following relation holds \cite{Gronau:2000zy,Gronau:2013mda}:
\beq\label{eq:cgammaratio}
\left[R_{C\Gamma}\right]_{U-{\rm spin}}=-1.
\eeq
Somewhat surprisingly, the experimental data show a strong violation of this relation:
\beq\label{eq:cgammaexp}
\left[R_{C\Gamma}\right]_{\rm exp}=-2.8\pm0.5.
\eeq
With our parametrization, and to first order in the $U$-spin breaking parameters
\beq\label{eq:cgammausb}
R_{C\Gamma}=-[1+2\hat t_r+2\hat p_r+2(\hat T_R/\hat T_i)\hat\Delta_i].
\eeq
We learn that the large deviation of $R_{C\Gamma}$ from the $U$-spin limit prediction is a consequence of (twice) three breaking parameters that add up in the same direction.

\section{Beyond the SM}
\label{sec:np}
It is not a simple task to establish, or to constrain, new physics contributions to CP asymmetries (and even more so to decay rates) in decay modes where there are two \ac{SM} contributions that differ in their weak and strong phases. Without any extra information, a \ac{SM} interpretation of the measurements when new physics actually plays a role is (almost) always possible.

In the case of $B_s\to K^+K^-$, there is, however, extra information, which is provided by its (approximate) $U$-spin relation with $B_d\to\pi^+\pi^-$ \cite{Barel:2020jvf,Fleischer:2010ib,Ciuchini:2012gd,Fleischer:2016jbf}. In the presence of new physics, a \ac{SM} analysis would lead to wrong values of $\hat P^q$ and $\hat T^q$. In particular, the new physics contributions might mimic $U$-spin breaking effects. Thus, the fact that a \ac{SM} analysis of the experimental data led to small $U$-spin breaking effects suggests that we can constrain the new physics by demanding that it does not lead to spurious $U$-spin breaking that is larger than observed. Such constraints assume that there is no cancellation between new physics effects and genuine $U$-spin breaking effects.

To see how new physics can mimic $U$-spin breaking, we now present our formalism for including new physics.
Any contributions to the four decay amplitudes of interest can be written as follows: 
\beqa
A_{KK}&=&\hat P^s\lambda^c_{bs}+\hat T^s\lambda^u_{bs}=\hat P^s\lambda^c_{bs}(1+r_s\lambda^u_{bs}/\lambda^c_{bs}),\no\\
\overline{A}_{KK}&=&\hat P^s\lambda^{c*}_{bs}+\hat T^s\lambda^{u*}_{bs}=\hat P^s\lambda^{c*}_{bs}(1+r_s\lambda^{u*}_{bs}/\lambda^{c*}_{bs}),\no\\
A_{\pi\pi}&=&\hat P^d\lambda^c_{bd}+\hat T^d\lambda^u_{bd}=\hat P^d\lambda^c_{bd}(1+r_d\lambda^u_{bd}/\lambda^c_{bd}),\no\\
\overline{A}_{\pi\pi}&=&\hat P^d\lambda^{c*}_{bd}+\hat T^d\lambda^{u*}_{bd}=\hat P^d\lambda^{c*}_{bd}(1+r_d\lambda^{u*}_{bd}/\lambda^{c*}_{bd}).
\eeqa
Distinguishing the \ac{SM} and new physics contributions, we write:
\beqa\label{eq:akkappnp}
A_{KK}&=&\hat P^s_{\rm SM}\lambda^c_{bs}(1+r^s_{\rm SM}R^{bs}_{uc}e^{+i\gamma}+r^s_{\rm NP}e^{+i\theta_s}),\no\\
A_{\pi\pi}&=&\hat P^d_{\rm SM}\lambda^c_{bd}(1-r^d_{\rm SM}R^{bd}_{uc}e^{+i\gamma}+r^d_{\rm NP}e^{+i\theta_d}),
\eeqa
where, in $\overline{A}_{KK}$ and $\overline{A}_{\pi\pi}$, $e^{i\gamma}$ and $e^{i\theta_q}$ are complex-conjugated.
Without loss of generality, we can rewrite these amplitudes as follows:
\beqa
A_{KK}&=&\hat P^s_{\rm SM}\lambda^c_{bs}\left[1+a_sr^s_{\rm NP}+R^{bs}_{uc}e^{+i\gamma}\left(r^s_{\rm SM}+b_s r^s_{\rm NP}\right)\right],\no\\
A_{\pi\pi}&=&\hat P^d_{\rm SM}\lambda^c_{bd}\left[1+a_dr^d_{\rm NP}-R^{bd}_{uc}e^{+i\gamma}\left(r^d_{\rm SM}+b_d r^d_{\rm NP}\right)\right],
\eeqa
where
\beqa\label{eq:asbsadbd}
a_s&=&\frac{\sin(\gamma-\theta_s)}{\sin\gamma},\ \ \ b_s=\frac{\sin\theta_s}{R^{bs}_{uc}\sin\gamma},\no\\
a_d&=&\frac{\sin(\gamma-\theta_d)}{\sin\gamma},\ \ \ b_d=-\frac{\sin\theta_d}{R^{bd}_{uc}\sin\gamma}.
\eeqa
Matching to the parametrization of Eqs. (\ref{eq:AkkpipiU}), we have
\beqa
\hat P^s=\hat P^s_{\rm SM}(1+a_s r^s_{\rm NP}),\ \ \ 
r_s&=&\frac{r^s_{\rm SM}+b_sr^s_{\rm NP}}{1+a_sr^s_{\rm NP}},\no\\
\hat P^d=\hat P^d_{\rm SM}(1+a_d r^d_{\rm NP}),\ \ \
r_d&=&\frac{r^d_{\rm SM}+b_dr^d_{\rm NP}}{1+a_dr^d_{\rm NP}}.
\eeqa

For our purposes, We assume that $|r^q_{\rm NP}|\ll1,|r^q_{\rm SM}|$, and expand the following two ratios to first order in $r^q_{\rm NP}$:
\beqa\label{eq:pspdrsrdnp}
\frac{\hat P^s}{\hat P^d}&=&\frac{\hat P^s_{\rm SM}}{\hat P^d_{\rm SM}}\left[1+a_s r^s_{\rm NP}-a_d r^d_{\rm NP}\right],\\
\frac{r_s}{r_d}&=&\frac{r^s_{\rm SM}}{r^d_{\rm SM}}\left[1+ 
(r^s_{\rm NP}/r^s_{\rm SM})b_s-(r^d_{\rm NP}/r^d_{\rm SM})b_d
-a_sr^s_{\rm NP}+a_dr^d_{\rm NP}\right].\no
\eeqa

Before proceeding, we note that by factoring out of the \ac{SM} contributions the CKM parameters $\lambda^c_{bq}$ and $R^{bq}_{uc}$, we leave in $\hat P^q_{\rm SM}$ and $r^q_{\rm SM}$ only QCD matrix elements and flavor-universal electroweak parameters, such as $G_F$. Thus, in the $U$-spin limit, we have $\hat P^s_{\rm SM}=\hat P^d_{\rm SM}$ and $r^s_{\rm SM}=r^d_{\rm SM}$. In contrast, $r^q_{\rm NP}$ does include in it, in general, flavor-dependent factors, and thus $U$-spin does not imply $r^s_{\rm NP}=r^d_{\rm NP}$.

In the $U$-spin limit, we thus obtain from Eqs. (\ref{eq:pspdrsrdnp}) 
\beqa\label{eq:pspdrsrdnpu}
\frac{\hat P^s}{\hat P^d}-1&=&a_s r^s_{\rm NP}-a_d r^d_{\rm NP},\\
\frac{r_s}{r_d}-1&=& (r^s_{\rm NP}/r_{\rm SM})b_s-(r^d_{\rm NP}/r_{\rm SM})b_d
-a_sr^s_{\rm NP}+a_dr^d_{\rm NP}.\no
\eeqa
We learn that new physics contributions can lead to $|\hat P^d/\hat P^d|\neq1$, and/or to $|r_s/r_d|\neq1$ even with no $U$-spin breaking. We recall that Eq.~(\ref{eq:exprsrdpspd}) gives, for the experimental central values, 
\beqa
|\hat P^s/\hat P^d|_{\rm exp}-1&\approx&0.26,\no\\
|r_s/r_d|_{\rm exp}-1&\approx&0.07.
\eeqa
In what follows we require that the new physics contributions do not lead to much larger deviations from unity.

We translate the upper bounds on the new physics contributions to lower bounds on the scale of new physics $\Lambda_{\rm NP}$. To do so, we assume that $\Lambda_{\rm NP}\gg m_W$, so that the new physics can be presented by dimension-six terms, 
\beq
\frac{g^2 X_{bq}}{\Lambda_{\rm NP}^2}(\overline{b_L}\gamma^\mu u_L)(\overline{u_L}\gamma_\mu q_L),
\eeq
and we include in $X_{bq}$ only flavor dependent factors. We will compare $g^2 X_{bq}/\Lambda_{\rm NP}^2$ to the size of the SM tree level contribution, which we take to be $g^2 \lambda^u_{bq}/m_W^2$, with $g$ the weak coupling constant. Concretely, we study five classes of models:
\begin{itemize}
\item Flavor anarchy: $X_{bq}={\cal O}(1)$.
\item Flavor anarchy with phase alignment: $X_{bq}=e^{i\gamma}{\cal O}(1)$.
\item \ac{FN} symmetry: $X_{bq}=e^{i\theta_q}{\cal O}(|\lambda^c_{bq}|)$.
\item \ac{GMFV}: $X_{bq}=e^{i\theta}\lambda^t_{bq}$.
\item \ac{MFV}: $X_{bq}=\lambda^t_{bq}$.
\end{itemize}
 
\subsection{Flavor anarchy} 
We refer to new physics as being anarchic when it has neither flavor suppression nor phase alignment. In other words, it is suppressed by its high scale (and possibly a loop factor) but by no other small parameters, $X_{bq}={\cal O}(1)$. In this case, we expect that
\beqa\label{eq:flaana}
r^s_{\rm NP}&=&r^s_{\rm AN}/|\lambda^c_{bs}|,\no\\ 
r^d_{\rm NP}&=&r^d_{\rm AN}/|\lambda^c_{bd}|,
\eeqa
where the CKM factors compensate for the $\lambda^c_{bs}$ and $\lambda^c_{bd}$ factors that are pulled out of the parenthesis in Eqs.~(\ref{eq:akkappnp}), and $r^s_{\rm AN}$ and $r^d_{\rm AN}$ are of the same order of magnitude but not equal. 

Examining Eqs.~(\ref{eq:pspdrsrdnpu}) and (\ref{eq:asbsadbd}), we learn that the largest modification will be due to the $b_s$ term, which is enhanced by $(R^{bs}_{uc})^{-1}$.  The physics behind this result is simple: the \ac{SM} contributions to $\hat P^s$, $\hat P^d$, $\hat T^s$ and $\hat T^d$ are proportional to $\lambda^c_{bs}$, $\lambda^c_{bd}$, $\lambda^u_{bs}$ and $\lambda^u_{bd}$, respectively. Anarchic new physics will modify most strongly the term that within the \ac{SM} is the most strongly CKM-suppressed, which is $\hat T^s$. 

In the $U$-spin limit, we obtain:
\beqa\label{eq:pspdrsrdan}
\frac{\hat P^s}{\hat P^d}&=&1+\frac{s_{\gamma-\theta_s}}{s_\gamma}\frac{r^s_{\rm AN}}{|\lambda^c_{bs}|}
                                                   -\frac{s_{\gamma-\theta_d}}{s_\gamma}\frac{r^d_{\rm AN}}{|\lambda^c_{bd}|},\\
\frac{\hat T_s}{\hat T_d}&=&1+ 
\frac{s_{\theta_s}}{s_\gamma}\frac{r^s_{\rm AN}}{r_{\rm SM}|\lambda^u_{bs}|}
+\frac{s_{\theta_d}}{s_\gamma}\frac{r^d_{\rm AN}}{r_{\rm SM}|\lambda^u_{bd}|},\no
\eeqa
where we use the notation $s_\gamma\equiv\sin\gamma$, and similarly to all phases.
The ratio between the corrections to $r_s/r_d$ and $\hat P_s/\hat P_d$ can be estimated as follows: 
\beq
\frac{|r_s/r_d|-1}{|\hat P^s/\hat P^d|-1}\approx\frac{1}{r_{\rm SM}}\frac{|\lambda^c_{bd}|}{|\lambda^u_{bs}|}\sim2.
\eeq
Since the corrections are comparable, the cancellation between the contributions to $\hat T^s/\hat T^d$ and $\hat P^s/\hat P^d$ could be accidental, so we require that $(b_s r^s_{\rm AN})/(r_{\rm SM}|\lambda^c_{bs}|)\lsim0.30$. Given that we consider anarchic new physics, we further assume that  $\sin\theta_s/\sin\gamma={\cal O}(1)$ and that $r_{\rm AN}/r_{\rm SM}\sim m_W^2/\Lambda_{\rm AN}^2$, where $\Lambda_{\rm AN}$ is the high scale of the anarchic new physics. We obtain the following bound:
\beq
\frac{r^s_{\rm AN}}{r_{\rm SM}|\lambda^u_{bs}|}\approx1200\frac{m_W^2}{\Lambda_{\rm AN}^2}\lsim0.30\ \Longrightarrow\ \Lambda_{\rm AN}\gsim  60m_W\sim 5\ {\rm TeV}.
\eeq
%

\subsection{Flavor anarchy with phase alignment}
The largest spurious $U$-spin breaking due to new physics occurs when the flavor structure is anarchic, as in Eq.~(\ref{eq:flaana}), but the new physics phase is set at a special value. Concretely, we consider the case where the new physics phases assume the values of
\beq
\theta_s=\theta_d=\gamma.
\eeq
The consequences of this scenario can be straightforwardly read from Eqs.~(\ref{eq:pspdrsrdan}):
\beqa\label{eq:pspdrsrdpa}
\frac{\hat P^s}{\hat P^d}&=&1,\no\\
\frac{\hat T_s}{\hat T_d}&=&1+\frac{r^s_{\rm PA}}{r_{\rm SM}|\lambda^u_{bs}|}
+\frac{r^d_{\rm PA}}{r_{\rm SM}|\lambda^u_{bd}|}.
\eeqa  
In this scenario, there is no deviation from the $U$-spin relation for $|\hat P^s/\hat P^d|$, but the $U$-spin relation for $|r_s/r_d|$ is violated. Thus, we require that the contribution to $|r_s/r_d|-1$ does not exceed the experimental value, for which  Eq.~(\ref{eq:exprsrdpspd}) gives $|r_s/r_d|_{\rm exp}-1\approx0.07$. Estimating for the flavor-anarchic phase-aligned case $r_{\rm PA}/r_{\rm SM}\sim m_W^2/\Lambda_{\rm PA}^2$, we require:
\beq
\frac{r_{\rm PA}}{r_{\rm SM}}\left(\frac{1}{|\lambda^u_{bs}|}+\frac{1}{|\lambda^u_{bd}|}\right)\approx1400\frac{m_W^2}{\Lambda_{\rm PA}^2}\lsim0.07\ \Longrightarrow\ 
\Lambda_{\rm PA}\gsim 140m_W\sim11\ {\rm TeV}.
\eeq
%

\subsection{The Froggatt-Nielsen mechanism}
In the \ac{FN} framework \cite{Froggatt:1978nt,Leurer:1992wg}, there is a $U(1)_{\rm FN}$ symmetry that is broken by a small spurion $\lambda_{\rm FN}$\,. The small flavor parameters -- mass ratios and CKM angles --  are accounted for by different powers of $\lambda_{\rm FN}$, depending on the charges of the relevant fields. The small parameter $\lambda_{\rm FN}$ is commonly taken to be of the order of the Cabibbo angle, $\lambda_{\rm FN}\approx0.2$, and conventionally taken to carry charge $Q_{\rm FN}(\lambda_{\rm FN})=-1$ under the $U(1)_{\rm FN}$ symmetry. For our purpose, however, where we aim to find the parametric suppression of flavor-changing dimension-six terms, one can relate this suppression directly to the suppression of the CKM angles (and, in some case, quark masses). Thus, if the new physics is subject to the \ac{FN} selection rules, then the leading contributions to $b\to u\bar u q$ are suppressed by $\lambda_{\rm FN}^{[Q_{\rm FN}(b_L)-Q_{\rm FN}(q_L)]}$, resulting in $X_{bq}=e^{i\theta_q}\times{\cal O}(|\lambda^c_{bq}|)$. Matching to Eqs.~(\ref{eq:akkappnp}), \ac{FN} implies:
\beqa
r^s_{\rm NP}e^{+i\theta_s}&=&r_{\rm FN}^se^{+i\theta_s},\no\\
r^d_{\rm NP}e^{+i\theta_d}&=&-r_{\rm FN}^de^{+i\theta_d}.
\eeqa

$U$-spin implies $\hat P^s_{\rm SM}=\hat P^d_{\rm SM}$ and $r^s_{\rm SM}=r^d_{\rm SM}$, {\it but not} $r^s_{\rm FN}=r^d_{\rm FN}$. The \ac{FN} selection rules imply that both $r^s_{\rm FN}$ and $r^d_{\rm FN}$ are ${\cal O}(1)$ (namely  not suppressed by powers of $\lambda_{\rm FN}$) but not equal. Using $U$-spin, we obtain:
\beqa
a_s&=&\frac{s_{\gamma-\theta_s}}{s_\gamma},\ \ \ b_s=\frac{s_{\theta_s}}{R^{bs}_{uc}s_\gamma},\no\\
a_d&=&\frac{s_{\gamma-\theta_d}}{s_\gamma},\ \ \ b_d=-\frac{s_{\theta_d}}{R^{bd}_{uc}s_\gamma},
\eeqa
and
\beqa\label{eq:usbfn}
\frac{\hat P^s}{\hat P^d}&=&1+\frac{s_{\gamma-\theta_s}}{s_\gamma} r^s_{\rm FN}+ \frac{s_{\gamma-\theta_d}}{s_\gamma}r^d_{\rm FN},\no\\
\frac{\hat T^s}{\hat T^d}&=&1+\frac{s_{\theta_s}}{s_\gamma}\frac{r^s_{\rm FN}}{r_{\rm SM}R^{bs}_{uc}}  - \frac{s_{\theta_d}}{s_\gamma}\frac{r^d_{\rm FN}}{r_{\rm SM}R^{bd}_{uc}}.
\eeqa

Assuming that $s_{\theta_q}={\cal O}(1)$, we can obtain a lower bound on the scale of new physics. The strongest bound can be obtained by noticing that \ac{FN} predicts a much larger deviation from the $U$-spin relations in $|r_s/r_d|$ than in $|\hat P_s/\hat P_d|$, in contrast to the experimental data (and to the expectations from naive factorization). The former is enhanced over the latter by a factor of order 
\beq
\frac{|r_s/r_d|-1}{|\hat P^s/\hat P^d|-1}\approx\frac{1}{r_{\rm SM}R^{bs}_{uc}}\sim10.
\eeq
Thus, we require that the \ac{FN} contribution to $|r_s/r_d|-1$ does not exceed $|r_s/r_d|_{\rm exp}-1\approx0.07$. Taking $s_{\theta_s}/s_\gamma={\cal O}(1)$, and estimating, for \ac{FN} models, $r_{\rm FN}/r_{\rm SM}\sim m_W^2/\Lambda_{\rm FN}^2$, we require:
\beq
\frac{r_{\rm FN}}{r_{\rm SM}}\frac{1}{R^{bs}_{uc}}\approx50\frac{m_W^2}{\Lambda_{\rm FN}^2}\lsim0.07\ \Longrightarrow\ 
\Lambda_{\rm FN}\gsim 26m_W\sim2\ {\rm TeV}.
\eeq
%

\subsection{General MFV}
In the \ac{GMFV} framework \cite{Kagan:2009bn}, the only sources of flavor $[U(3)]^5$ breaking are the Yukawa matrices of the \ac{SM}, but there could be new sources of (flavor-universal) CP violation, $X_{bq}=e^{i\theta}\lambda^t_{bq}$. (For a study in a related framework, see ref.~\cite{Crivellin:2019isj}.)
Matching to Eqs.~(\ref{eq:akkappnp}), \ac{GMFV} implies:
\beqa
r^s_{\rm NP}e^{+i\theta_s}&=&-r_{\rm GMFV}^se^{+i\theta}(1+R^{bs}_{uc}e^{i\gamma}),\no\\
r^d_{\rm NP}e^{+i\theta_d}&=&-r_{\rm GMFV}^de^{+i\theta}(1-R^{bd}_{uc}e^{i\gamma}).
\eeqa
$U$-spin implies $r_{\rm GMFV}^s=r_{\rm GMFV}^d$. With these replacements, and using $U$-spin, we obtain:
\beqa\label{eq:abgmfv}
&&a_s=\frac{s_{\gamma-\theta}-R^{bs}_{uc}s_\theta}{s_\gamma},\ \ \ 
b_sR^{bs}_{uc}=\frac{s_\theta+R^{bs}_{uc}s_{\theta+\gamma}}{s_\gamma}\no\\
&&a_d=\frac{s_{\gamma-\theta}+R^{bd}_{uc}s_\theta}{s_\gamma},\ \ \ 
-b_dR^{bd}_{uc}=\frac{s_\theta-R^{bd}_{uc}s_{\theta+\gamma}}{s_\gamma}.
\eeqa
and
\beqa\label{eq:usbgmfv}
\frac{\hat P^s}{\hat P^d}&=&1+r_{\rm GMFV}\frac{s_\theta}{s_\gamma}(R^{bd}_{uc}+R^{bs}_{uc}),\no\\
\frac{\hat T^s}{\hat T^d}&=&1-\frac{r_{\rm GMFV}}{r_{\rm SM}}\frac{s_\theta}{s_\gamma}\left(\frac{1}{R^{bs}_{uc}}+\frac{1}{R^{bd}_{uc}}\right).
\eeqa

Assuming that $s_\theta={\cal O}(1)$, we can obtain a lower bound on the scale of new physics. The strongest bound can be obtained by noticing that \ac{GMFV} predicts a much larger deviation from the $U$-spin relations in $|r_s/r_d|$ than in $|P_s/P_d|$:
\beq
\frac{|r_s/r_d|-1}{|\hat P^s/\hat P^d|-1}\approx\frac{1}{r_{\rm SM}}\frac{1}{R^{bd}_{uc}R^{bs}_{uc}}\sim25.
\eeq
Thus, we require that the \ac{GMFV} contribution to $|r_s/r_d|-1$ does not exceed $|r_s/r_d|_{\rm exp}-1\approx0.07$. Taking $s_\theta/s_\gamma={\cal O}(1)$, and estimating, for \ac{GMFV} models, $r_{\rm GMFV}/r_{\rm SM}\sim m_W^2/\Lambda_{\rm GMFV}^2$, we require:
\beq
\frac{r_{\rm GMFV}}{r_{\rm SM}}\frac{1}{R^{bs}_{uc}}\approx50\frac{m_W^2}{\Lambda_{\rm GMFV}^2}\lsim0.07\ \Longrightarrow\ 
\Lambda_{\rm GMFV}\gsim 26m_W\sim2\ {\rm TeV}.
\eeq
%

\subsection{Minimal flavor violation (MFV)}
In the \ac{MFV} framework \cite{DAmbrosio:2002vsn}, the only sources of flavor $[U(3)]^5$ breaking and of CP violation are the Yukawa matrices of the \ac{SM}, $X_{bq}=\lambda^t_{bq}$. \ac{MFV} implies:
\beqa
r^s_{\rm NP}e^{i\theta_s}&=&-r_{\rm MFV}(1+R^{bs}_{uc}e^{i\gamma}),\no\\
r^d_{\rm NP}e^{i\theta_d}&=&-r_{\rm MFV}(1+R^{bd}_{uc}e^{i\gamma})\,.
\eeqa

The consequences of \ac{MFV} can be straightforwardly derived by using the analysis of \ac{GMFV} with $\theta=0$. Eq.~(\ref{eq:abgmfv}) implies that, in this case,
\beq\label{eq:abmfv}
a_s=b_s=a_d=b_d=1\,,
\eeq
and Eq.~(\ref{eq:usbgmfv}) implies that, in the $U$-spin limit,
\beq
\hat P_s/\hat P_d=1,\ \ \ r_s/r_d=1.
\eeq
Thus, \ac{MFV} physics does not mimic $U$-spin breaking.

This result can be easily understood. For \ac{MFV} physics, the new physics contribution can be absorbed into $P_t$ of Eqs.~(\ref{eq:tsps}) and (\ref{eq:tdpd}), and thus the \ac{MFV} analysis cannot be distinguished from the \ac{SM} analysis. 

While the constraints from $U$-spin breaking cannot be applied to the \ac{MFV} case, one can still bound the size of the \ac{MFV} contribution. Assuming $U$-spin, we have
\beq
\hat T=T+P_u-P_t-P_{\rm MFV},\ \ \ \hat P=P_c-P_t-P_{\rm MFV}.
\eeq
Approximating $P_u\approx P_c$, we obtain
\beq 
r={\hat T}/{\hat P}\approx{T}/{\hat P}+1.
\eeq
Obviously, when $|T/\hat P|\ll1$, we have $r\simeq1$, inconsistent with the experimental result that $r\approx4.8$. We learn that we can require that $|T/\hat P|\gsim3.8$, which translates into
\beq
\Lambda_{\rm MFV} \gsim 2m_W\approx 160\ {\rm GeV}.
\eeq
%

\acresetall
\section{Conclusions}
The recent measurements of two CP asymmetries in $B_s\to K^+K^-$, $S_{KK}$ and $C_{KK}$, constitute the first observation of time-dependent CP violation in the neutral $B_s$ system. While the theoretical analysis of these observables suffers from hadronic uncertainties, the approximate $U$-spin symmetry of QCD relates them to the corresponding observables in $B_d\to\pi^+\pi^-$, $S_{\pi\pi}$ and $C_{\pi\pi}$. The set of five observables - the four CP asymmetries and the ratio of decay rates, $\Gamma(B_s\to K^+K^-)/\Gamma(B_d\to\pi^+\pi^-)$, provides a new arena to test the \ac{SM}, the approximate $U$-spin symmetry, and the presence of flavor changing new physics. 

We find that the measured values are consistent with $U$-spin breaking at the level of 30\% or smaller. Furthermore, the $U$-spin breaking is a factor $\sim4$ smaller in the double ratio of matrix elements, $(\hat T_s/\hat T_d)/(\hat P_s/\hat P_d)$, than in each of these ratios separately. This result is consistent with the assumption that the leading effect is coming from $f_K/f_\pi$ which cancels in the factorizable contributions to the double ratio, with the remaining non-factorizable contributions and contributions of ${\cal O}(m_s/m_b)$ much smaller.

With regard to new physics, the main tool that we use to probe it is the observation that, in general, when interpreting the experimental results assuming the \ac{SM}, the new physics contributions will mimic $U$-spin breaking effects. If the new physics contributions are large, they can generate large spurious $U$-spin breaking. Assuming that there are no cancellations between the new physics effects and genuine $U$-spin breaking effects, we found lower bounds on the scale of classes of new physics with various flavor structures:  for flavor anarchy $\Lambda_{\rm AN}\gsim5$ TeV, for \ac{FN} selection rules $\Lambda_{\rm FN}\gsim2$ TeV, and for \ac{GMFV} $\Lambda_{\rm GMFB}\gsim2$ TeV.  We further demonstrated that new physics at a scale as high as ${\cal O}(10\ {\rm TeV})$ can generate significant spurious $U$-spin breaking.

New physics subject to \ac{MFV} does not mimic $U$-spin breaking. In fact, analyzing the results in the \ac{MFV} framework is identical to carrying out the analysis in the \ac{SM} framework. Yet, if the scale of \ac{MFV} new physics is below 160 GeV, it will bring the situation closer to a single phase dominance, suppressing the CP asymmetries to below their observed values, and is thus excluded. 

Our bounds on the scale of new physics assume that there are no cancellations between the spurious and the genuine $U$-spin breaking effects. Such cancellations can relax our bounds by factors of ${\cal O}(1)$. We further did not include loop factors for the new physics contributions. Note, however, that the relevant decay processes, $b\to u\bar uq$ ($q=s,d$), are not \ac{FCNC} processes, and have tree level contributions already in the \ac{SM}.  Finally, while $S_{KK}$ and $S_{\pi\pi}$ depend on, respectively, $B_s-\overline{B_s}$ and $B_d-\overline{B_d}$ mixing, which are \ac{FCNC} processes, we used purely experimental data to include the neutral meson mixing parameters, and thus our analysis is independent of the mixing mechanism.

\subsection*{Acknowledgements}
We are grateful to Yasmine Amhis, Yuval Grossman and Zoltan Ligeti for helpful discussions. YN is the Amos de-Shalit chair of theoretical physics, and is supported by grants from the Israel Science Foundation (grant number 1124/20), the United States-Israel Binational Science Foundation (BSF), Jerusalem, Israel (grant number 2018257), by the Minerva Foundation (with funding from the Federal Ministry for Education and Research), and by the Yeda-Sela (YeS) Center for Basic Research. IS~is supported by a fellowship from the Ariane de Rothschild Women Doctoral Program.


\end{document}